\documentstyle[preprint,aps]{revtex}
\tightenlines

\draft

\begin{document}
\title{ Comparison of tunneling through molecules with Mott-Hubbard and
with dimerization gaps.}
\vskip0.5truecm 
\author {Julien Favand and Fr\'ed\'eric Mila} 
\address{
      Laboratoire de Physique Quantique, Universit\'e Paul Sabatier\\
      31062 Toulouse (France)\\
      }
\maketitle

\vskip2.truecm 

\begin{abstract}
In order to study the tunneling of electrons through an interacting, 1D, 
dimerized
molecule connected to
leads, we consider the persistent current in a ring embedding this molecule.
We find numerically that, for spinless fermions, a molecule with a
gap mostly due to interactions, i.e. a Mott-Hubbard gap, gives rise to a 
larger persistent current than a molecule with the same gap, but
due only to the dimerization. In both cases, the
tunneling current decreases exponentially with the size of the molecule, but
more slowly in the interacting case. Implications for molecular electronic are
briefly discussed.
\end{abstract}

\vskip1.truecm 


\section{Introduction}

One of the major challenges of molecular electronics is to find molecules
which could act as wires. For this purpose,
special attention has been devoted to conjugated oligomers, and especially to 
the
simplest of them, polyene, the finite size equivalent of polyacetylene
($C_pH_{p+2}$). 
If one considers non-interacting particles, the
conductance of the set lead-molecule-lead is, at low bias and
according to Landauer's formula \cite{Landauer,Stone},
proportional to the 
transmission coefficient through the
molecule at the Fermi level of the leads.
This approach was 
already successful
in reproducing experimental results on a single $C_{60} $ molecule 
\cite{Joachim1}, and it 
seems natural to apply it to polyene.

In the framework of non-interacting electrons, 
there is a charge gap
due to the dimerization of the chain and Joachim 
and collaborators \cite{Joachim2} found that
the conductance shows a minimum when the
Fermi energy of the leads is in the middle of this gap.
At this point and only here, the current changes linearly with the voltage.
Such a property is important to get a good molecular wire. Thus we have to
consider the magnitude of the current at this point.
Joachim showed that this minimum
behaves as $t_0e^{-\gamma (2N-2)}$, $2N$ being the length of the molecule. 
The coefficient $\gamma$
grows with the dimerization, and the prefactor $t_0$
depends on the lead-molecule
contact. It turns out to be possible to tune these parameters to get 
non negligible currents in
long wires (10 nm).
These preliminary results were then confirmed by further simulations
on polyenes and alkanes connected to golden
leads. See also Ref.\cite{Magoga}.

However, there are very good reasons to believe that
electronic correlations play a role in fixing the magnitude of the
charge gap in polyene \cite{Baeriswyl1}. 
We give here the main ones.
First, the lowest excited state is dipole-allowed, and this 
can be recovered theoretically only when correlations are included 
\cite{Soos}. Second, the observed
negative spin densities 
\cite{Thomann,Kahol} can be understood only for models with
interaction \cite{Hirsch}. Third, photo-induced absorption experiments
\cite{Friend}
showing a splitting of the soliton peak in the middle of the gap would be
puzzling without interactions \cite{Baeriswyl2}. Furthermore, ab-initio
calculations need to include in some way the correlation to get the right
ground state dimerization \cite{Fulde}.
Therefore, we wish to include interactions inside the molecule and treat them
exactly.
In comparison with the free case and for a given gap, we would like to know if 
these correlations increase or decrease the lowest current.

How can we study this current?
Interesting results have been derived for a Luttinger liquid connected to leads
\cite{Safi,Maslov},
but they cannot be applied to the present case when the chemical potential 
lies in the gap.
There is also a general formula describing the current through an interacting
region connected to non-interacting leads 
\cite{Meir}. 
However this formulation only allows an analytical evaluation if the 
interactions
can be treated perturbatively, which is clearly not the case in polyene,
and it is not so clear how to use numerical results obtained on finite size
clusters for the Green's functions within this formalism.

So, we need another way to estimate the conductance. In the non-interacting 
case,
the behavior of the conductance is strongly related to the exponential decrease
of the amplitude of the wave function at the Fermi level inside the molecule.
( We consider only one channel for simplicity. )
This exponential decrease also controls the persistent current in a ring
embedding the molecule. So, persistent currents and conductance are related
in this case. 
We will assume that such a relationship still holds, at least qualitatively, 
in the interacting case.
More precisely, we will suppose that what happens to the minimum persistent 
current by changing the molecular dimerization or interactions informs us on
what would happen to the tunneling current, and we will concentrate on the
persistent current because we know how to evaluate it numerically.

The paper is organized as follows. We introduce the model in section II.
In section III, we analyze in more details the relationship between the 
persistent current
and the conductance in the non-interacting case. Finally we present the central
results of this paper about the 
persistent currents in the interacting case in section IV.

\section{The model.}

We describe the molecule by
a dimerized and interacting region of length $M$.
It is embedded
in a non-interacting ring of length $N$ describing the electrode (see Figure1).
The ring is pierced by an  Aharonov-Bohm flux $\Phi$.
Denoting by $L=N+M$ the total number of sites, the persistent 
current $I(\Phi_0)$ is
related to the groundstate energy $E(\Phi)$ by
$I(\Phi_0)= L \frac{dE}{d\Phi}\huge|_{\phi=\phi_0}$. 
for simplicity we
will concentrate on the mean 
value of
$|I|$ over $\Phi_0$ given by $\Delta I =\frac{L}{\pi} |E(\pi)-E(0)|$.

We will use the following hamiltonian :
\begin{equation}
\displaystyle\hat H^ = \displaystyle\hat H^{e}
+ \displaystyle\hat H^{\alpha}
+ \displaystyle\hat H^{m}
+ \displaystyle\hat V
+ \displaystyle\hat \epsilon_m
\end{equation}

with

\begin{eqnarray}
\displaystyle\hat H^{e} & = & 
-t e^{i\phi}\sum^{N-2}_{i=0}c{}^\dagger_{i+1}c{}^{\ }_i +h.c. \nonumber \\
\displaystyle\hat H^{\alpha} & = &  
(-\alpha e^{-i\phi} c{}^\dagger_{L-1}c{}^{\ }_0 
-\alpha e^{i\phi} c{}^\dagger_{N}c{}^{\ }_{N-1}) ) +h.c. \nonumber  \\
\displaystyle\hat H^{m} & = & 
-t_1 e^{i\phi} \sum^{M/2-1}_{i=0}c{}^\dagger_{N+2i+1}c{}^{\ }_{N+2i} +h.c.
-t_2 e^{i\phi}\sum^{M/2-1}_{i=1}c{}^\dagger_{N+2i}c{}^{\ }_{N+2i-1}+h.c.
\nonumber \\
\displaystyle\hat V & = & 
V \sum^{L-2}_{i=N} n_{i} n_{i+1} \nonumber   \\
\displaystyle\hat \epsilon_m & = & 
\epsilon_m \sum^{L-1}_{i=N} n_i \nonumber 
\end{eqnarray}

In these expressions, 
the operator $c{}^\dagger_{j}$ creates a fermion at site $j$, and 
$n_i=c{}^\dagger_{i}c{}_{i}$ is the density operator at site i.
The metallic electrode is described by a tight-binding hamiltonian 
$\displaystyle\hat H^e$ 
with a hopping integral $t$.

The molecule is described by the sum of three terms: 
$\displaystyle\hat H^m$ is a tight-binding
hamiltonian that describes the kinetic energy inside the molecule. It involves
two
alternating
hopping integrals $t_1$ and $t_2$ to take the dimerization into account.
$\displaystyle\hat V$ represents the nearest
neighbour repulsion between the particles, and $\hat \epsilon_m$ 
fixes the chemical potential 
of the molecule.

The molecule and the electrode are connected by
a transfer term $\hat H^{\alpha}$ that allows the electrons to hop from one to
the other.

Finally all the hopping integrals are multiplied by a phase factor $e^{i\phi}$
with $\phi=\Phi/L$ to describe
the Aharonov-Bohm flux.

Of course a realistic calculation should deal with electrons, i.e. fermions with
spin. It turns out however that the sizes one can reach with electrons are too
small to allow a meaningful finite size analysis (see below). So we have 
decided to restrict ourselves to spinless fermions. This should be useful as
a first step toward more realistic systems because the physics of 
Mott-Hubbard insulators - i.e. systems where the charge gap is due to
correlations - is very similar for spinless fermions and fermions with
spin.

\section{Relationship between the persistent current and the conductance
for non-interacting particles}

There is no exact relation between the persistent
current and the conductance in general, but we will show that the two quantities
share qualitative and quantitative features in the non-interacting case.

\subsection{How to calculate the conductance G?}

According to Landauer, the conductance is proportional to the transmission
coefficient $T_F$ at the Fermi level. This coefficient can be 
calculated by considering
two semi infinite leads connected to the molecule. Using 
the matching procedure, the stationary state $|E>$ of energy $E$
can be searched for as:
\begin{equation}
|E>=|k^{e}_{L}>+r_E|-k^{e}_{L}>+A^{m}_1|k^{m}>+A^{m}_2|-k^{m}>+t_E|k^{e}_{R}> 
\end{equation}
where $r_E$, $A^{m}_1$, $A^{m}_2$ and $t_E$ are unkonwn coefficients to be
determined by continuity conditions.
Here $|k^{e}_{L,R}>$ and $|k^{m}>$ correspond to the states with energy $E$ 
that one would get for infinite leads or for an infinite molecule, and 
the momenta must satisfy:
$E=-2t\cos(k^{e}_j)=-\sqrt{t_1^2+t_2^2+2t_1t_2\cos(2 k^{m}_j})$.
The transmission coefficient is then obtained as $T(E)=|t_E|^2$.
An analytical formula for $T(E)$ can be derived, but is is too complicated 
to be written down. It is also possible to get $T(E)$ for more realistic band
structures. See Ref.\cite{Ratner}.
A typical example is depicted in Figure 2. Although the method is different from
that of Ref.\cite{Joachim2}, we have checked that the results are indeed the
same.

\subsection{How to calculate the persistent current $\Delta I$ ?}

We now consider the geometry of Figure 1. 
In the absence of correlations,
the total energy $E$ inside the ring is the sum of the energies $E_i$ of the 
occupied
monoparticular states (up to $E_F$). Thus
the total current $I$ is the sum of the individual currents $I_i$ of these 
levels.

The simplest way to get these energies $E_i$ is again to use the 
matching procedure.
The ring being closed, a monoparticular state $|E_i>$ must now be searched for
as:

\begin{equation}
|E_i>=A^{e}_1|k^{e}_1>+A^{e}_2|k^{e}_2>+A^{m}_1|k^{m}_1>+A^{m}_2|k^{m}_2>
\end{equation}
where $|k^{e}_j>$ and $|k^{m}_j>$ are again free propagating waves
in the electrode and in the molecule. Note however that they now satisfy:
$E_i=-2t\cos(k^{e}_j-\phi)=-\sqrt{t_1^2+t_2^2+2t_1t_2\cos(2 k^{m}_j-2 \phi)}$,
$j = {1,2}$, because of the Aharonov--Bohm flux.

The continuity conditions yield 4 linear equations for the coeficients
$A^{e}_i$
and $A^{m}_i$. We get a solution $E_i$
each time the determinant of this system vanishes. This determinant can be
evaluated numerically, and we get the monoparticular states and 
energies as in the preceding subsection.

An alternative way consists in using the transfer matrix formalism. 
If $\alpha$
and $\beta$ are
the amplitudes of the incoming and outgoing waves on the left side of the 
molecule,
the incoming and outgoing amplitudes on the right side $\alpha '$ and $\beta '$ 
are given by
$$
\left(\matrix{\beta '\cr
            \alpha '\cr}\right)=\left(\matrix{1/t^{*}_k & -r^{*}_k/t^{*}_k \cr
                                                    -r_k/t_k & 1/t_k \cr}\right)
\left(\matrix{\alpha \cr
               \beta \cr}\right)
$$
where $t_k$ and $r_k$ the transmission and reflection coefficients
for an incident wave vector $k$ defined in the preceding subsection. The matrix
$T_m$ that enters this equation is called the transfer matrix of the molecule.

A similar definition for the transfer matrix $T_e$ of the electrode holds,
with $r_k=0$ and $t_k=e^{ikN}$.
The continuity conditions can then be written:

\begin{equation}
T_e \times T_m \left(\matrix{\alpha \cr
               \beta \cr}\right)
= e^{i\Phi}\left(\matrix{\alpha \cr
               \beta \cr}\right)
\end{equation}

Note that in this approach the flux is concentrated in the boundary conditions,
so that the energy is related to the wave--vectors by the usual relations of
subsection II-A.

Denoting by $\Theta_k$ the phase of $t_k$, 
the condition that the corresponding determinant vanishes yields: 
\begin{equation}
|t_k| \cos(\Phi)=\cos(\Theta_k +kN)
\label{valprop}
\end{equation}

This equation can be solved numerically for $k$, 
$t_k$ being calculated as in subsection A, and we get again the spectrum.
While the matching procedure is more convenient for that purpose, 
this equation turns out to be very useful to discuss the minimum 
value of the
persistent current as a function of the chemical potential $\epsilon_m$ 
(see subsection
II-D).

\subsection{General shape of G and $\Delta I$.}

We are now in a position to compare the behaviors of $\Delta I$ and $T$ with the
Fermi level $E_F$ of the electrode. More precisely, the important parameter
is the difference $E'=E_F-\epsilon_m$ which vanishes when $E_F$ sits
right in the middle of the molecular gap.

We can first control $E'$ by varying the value of $\epsilon_m$ for a given
filling of the ring.
The plot of $\Delta I(E')$ for a half-filled system is given in Figure 3.
Although their shape and size are modified, it is important to notice that the
resonances are located at the same energies as for $T(E_F)$.
Besides, as long as $\alpha$ is not too large - which is the case in Figure 3 -
the number of resonances is equal to 
$M$ for both. 
Some of them are 
washed out - both for $T$ and for $\Delta I$ - when
$\alpha$ is of the order of $t$ however.

We can also vary the filling $N_F$ of the ring (which control $E_F$) 
for a given 
$\epsilon_m$, say 0, and we
get  a similar result for $\Delta I$ as a function of $E'$
(See Figure 3.). The only difference is that some resonances are now missing
since the 
bandwidth of the electrode is smaller than that
of the molecule in the example of Figure 3. 
Of course they can all be restored by tuning the chemical potential of
the molecule
$\epsilon_m$.

\subsection{Value of the minimum current}

We now compare the minimum value of the persistent current and the minimum value
of the transmission coefficient. They both appear for $E_F=0$, if the
chemical potential of the molecule $\epsilon_m$ is set to 0.
Let us first calculate the mean current of the monoparticular state located at
the Fermi level 
$\displaystyle \Delta I_F=\frac{L}{\pi} | E_{k_F}(\Phi=\pi)-
E_{k_F}(\Phi=0) | $. The energy at the Fermi level is given by:

\begin{equation}
E_F(\Phi)=-2t \cos[k_F(\Phi)]
\label{courant}
\end{equation}

For a finite-size system, $E_F<0$ and $|E_F|<<1$, and Eq.(\ref{courant}) gives:

\begin{equation}
E_F(\Phi)=2t k_F(\Phi) -t\pi 
\label{Efeq}
\end{equation}

We also know that $ |t_F(\Phi)| << 1 $ and thus Eq. (\ref{valprop}) yield:

\begin{equation}
N k_F(\Phi) + \Theta_F(\Phi) = 2n \pi+ \epsilon \pi /2 
-\epsilon \cos \Phi |t_F(\Phi)|
\label{kfeq} 
\end{equation}

where $\epsilon =\pm 1$ and $n \in Z$ depend on N and the scattering properties
of the molecule, but do not depent on $\phi$ since
$k N + \Theta_k $ is a continuous function of $\phi$

We get from Eq. (\ref{Efeq}) and Eq. (\ref{kfeq}) :

\begin{equation}
E_F(\Phi)=-2t[ \Theta_F(\Phi) + 2n \pi+ \epsilon \pi /2 
-\epsilon \cos(\Phi)|t_F\Phi)| ) /N -t\pi ]
\end{equation}

Thus from Eq.(\ref{courant}) the current at the Fermi level is :

\begin{equation}
\Delta I_F = 2t |\epsilon(|t_F(\pi)|+
|t_F(0)|)-\Theta_F(\pi)+\Theta_F(0)|L/N 
\end{equation}

When N goes to infinity, N/L tends to 1, $|t_F(\pi)|$ and $|t_F(0)|$ merge
and $(\Theta_F(\pi)-\Theta_F(0))$ goes
to zero.

Finally :

\begin{equation}
\Delta I_F = 4t \sqrt{T_F}
\end{equation}

It is possible to see in the same way that the sign of the current carried by
the successive levels
is alternate. The absolute value of this currents changes continuously with the
energy level, and
therefore at large $L$, the total current, which is the sum of the currents of
the occupied level, is half of the last one. Thus, taking the absolute value : 
$\Delta I =\Delta I_F /2 $.

So, by using the Landauer formula $G=\displaystyle \frac{e^2}{h} T_F$, 
we get the following relation
between the persistent current and the conductance :

$$ \Delta I= \frac{2t}{e} \sqrt{hG} $$

The amplitude of a transmitted wave after the molecule is proportional to
$e^{-2\gamma M}$ with $2\cosh{\gamma} =t_1/t_2$.		
Thus $T_F$ and G decrease like $e^{-2 \gamma M}$ while 
$\Delta I$ is proportional to $e^{- \gamma M}$.
This point can be easily understood if we remember that we are free, by 
a gauge transformation, to
concentrate the whole phase factor $\Phi=\pi$ in the middle of the molecule.
The difference between the zero flux case and the $\Phi=\pi$ case in the
hamiltonian is $\delta \hat H= 2 \times t_2(c{}^\dagger_{\frac{M}{2}}
c{}^{\ }_{\frac{M}{2}-1} +hc)$.
This can be
seen as a perturbation and its first order influence on the level $E_i$ is :
$$\delta E_i = <\Psi_i | \delta \hat H | \Psi_i> = 2t_2[\Psi_i (M/2) \times 
\Psi{}^{*}_i (M/2-1)
+cc] $$
At the Fermi level we have :
$\Psi_f (M/2) \propto e^{-\gamma M/2}$ and thus 
$\delta E_F \propto e^{- \gamma M} $. This energy shift is proportional to the
Fermi-level current and thus to the total current.

\vskip0.4truecm 

As a conclusion,
in the non-interacting case, the persistent current gives valuable
informations on the conductance. First, the resonances have the same location. 
Second, the minimum persistent current is proportional to the square root of the
minimum conductance. Thus, since the later decreases exponentially with the size
of the molecule, the former has the same behavior, but with a coefficient in the
exponential
twice as small.

\section{Persistent current for an interacting molecule}

We now study the minimum persistent current in the ring when
the interaction term $\hat V$ is included.
We have used three different methods to
get the groundstate energy of the interacting system.
The last two are mainly used to check the reliability of the first one.

\subsection{Exact diagonalizations.}

We haved use Lanczos algorithm to diagonalize the hamiltonian and get the
groundstate energy $E_0(M,N)$ for small values of M and N. 
For a given molecule of length $M$, we wish to get rid 
of the effect of the finite size $N$ of the electrode.
This was achieved by performing a finite size scaling on $E(L)$
with fixed M and increasing N.  

It turns out that a meaningful description fo the molecule requires at least
$M=6$. Besides, to perform this scaling, we need as many values of 
$N$ as possible, and this can be conveniently done while keeping the density
fixed only at half--filling.
Then for 
fermions with
spin $\frac{1}{2}$ and for $L=16$ the dimension of the Hilbert space is   
of the order of $1.65\  10^8$, i.e. too large to be handled numerically.
Thus,
in order to get enough points to perform a reliable finite size scaling, 
we have decided to limit our 
study to interacting spinless fermions.
In that case, we could go up to $L=26$, in which case the dimension of the
Hilbert space is of the order of
$10^7$. 

If we plot the average persistent current versus $1/L$, is appears that 
a quadratic law 
fits the numerical results quite well (see Figure 4)
as long as the bandwidth of the electrode
is smaller than the molecule's one. This case corresponds to a 
large enough density of
states at the Fermi-energy of the electrode.
To test the reliability of the fit,
we have made the same plot
up to $L=120$ for non-interacting particles (see also Figure 4). The 
difference between the two extrapolated values for the two fits 
(up to $L=26$ and 
up to $L=120$)
is never
more than $5\%$.
However, this is no proof that the same fit is accurate for interacting 
particles, and we now turn to alternative methods to check the
extrapolation.

\subsection{Small repulsion limit.}

We will now derive the persistent current to first order in the interaction $V$.
Although this limit is not relevant for the case of polyene, it will be used 
to test the relibiality of our scaling law. The interaction term in the 
Hamiltonian is :

$$ \hat{V}= V \sum^{M-2}_{i=0}c{}^\dagger_{i}c{}^{\ }_{i} 
c{}^\dagger_{i+1}c{}^{\ }_{i+1} $$

The energy levels $|E_j>$ of the non-interacting case can be calculated as in
Section IB. They are non-degenerate and form a basis for the
monoparticular states. If the $Ej$'s are sorted in increasing order, 
the non-interacting ground state is given by:
$$ |\Phi^{0}>= \prod_{j\le N_F} c{}^\dagger_{E_j} | \emptyset > $$
where $c{}^\dagger_{E_j}$ creates one particle in state
$|E_j>$. Besides, at half--filling, $N_F=L/2$. 
The original creation operator $c^\dagger_i$ can then be written:
$$ c{}^\dagger_{i}= \sum^{N}_{j=1} c{}^\dagger_{E_j}<E_j|i> $$
The first order correction to the ground state energy 
$\Delta E=<\Phi^{0}|\hat{V}|\Phi^{0}>$ 
is thus given by:
\begin{equation}
\Delta E= V \sum^{M-2}_{i=0} ( \sum_{j \le N_F,l \le N_F} 
|<i+1|E_j>|^2 |<i|E_l>|^2 
\end{equation}
$$+ \sum_{j\le N_F,l>N_F}
<E_j|i+1><i+1|E_l><E_l|i><i|E_j> ) $$

$\Delta E$
can be easily calculated numerically for periodic ($\phi=0$) or antiperiodic
($\phi=\pi$) boundary conditions up to quite large systems. 
We can see on Figure 4 that the quadratic scaling law is
still valid up to $L=36$ for small $V$.

\subsection{DMRG.}

To find the ground state energy of a one-dimensional system, 
an alternative method to the exact diagonalisation is the Density Matrix
Renormalisation Group \cite{White}. While this method gives very precise 
results for open boundary conditions, it is always less accurate for 
closed boundary conditions. 

Treating the molecule exactly and considering half of the electrode
as the growing block, we meet strong limitations. Indeed, if $M=6$, adding 1
site at each junction between the molecule and the leads
to avoid artefacts and 2 sites to let the system grow, we have to treat 10 sites
exactly.
If we keep $m=100$ states
to describe this growing block, the dimension d of the global Hilbert space is
roughly : $d=m \times 2^{10} \times m =10^7 $
and it is difficult to do much better.

Therefore, we have only been able to get accurate results up to
$L=30$. We did not go further, because for $L=32$ and the maximum available 
m (actually 120), the
relative error on the current was already 5\% in the non-interacting case.
It is nevertheless satisfactory to see that the results are again 
consistent with the extrapolation of the exact 
diagonalization data, as shown in Figure 4.

\subsection{Results.}

\subsubsection{Variation of $\Delta I$ with $\epsilon_m$.}

The plot of $\Delta I$ versus $\epsilon_m$ (Figure 5) for a half-filled ring 
has the same shape
as in the non-interacting case (Figure 3). The main effect of the interaction
is to shift the location $\epsilon^{min}_m$ and the value $I_{min}$ of the
minimum current. But the 6 resonances are still present, although their
relative distances change with respect to the non-interacting case. 
The current $I_{min}$ is the one we are interested in.
And the corresponding $\epsilon^{min}_m$ does not change with $L$, but
depends on $V$ (it is roughly equal to
the opposite of $V$): $\epsilon^{min}_m$ counterbalances the mean
field effect of $V$.

\subsubsection{Effect of $V$ on $I$ for a given gap.}

In order to study the effect of electron-electron interaction on the minimum
current, we have to find a meaningful way to compare it to the non-interacting
case. Now, in a realistic situation, the rough value of the hopping 
integrals $t_1$ and $t_2$ is usually known - it is basically given by the total
bandwidth $2(t_1+t_2)$ - but the precise value of their ratio,
which together with the interaction term controls the charge gap $\Delta_c$ ,
is not known 
as
accurately. The gap itself is known quite accurately however. So we have decided
to compare models with the same values of $2(t_1+t_2)$ and $\Delta_c$. In a
given class of models, $V$ is then a function of $t_2/t_1$.

The main result of this paper is that,
for a given bandwidth and a given charge gap $\Delta_c$, the minimum persistent 
current $I_{min}$
is larger when the
interaction is partially responsible for the charge gap than when 
the charge gap is due only to the
dimerization.
If we plot $I_{min}$ versus $V$ for fixed $\Delta_c$ and bandwidth,
as in Figure 6 , this 
effect appears clearly, and
it is quite substantial. For example, if $2(t_1+t_2)=4t$ and $\Delta_c=3.02t$ 
we get
a current twice as large for $V=t$ than in the non-interacting
case.
Note on Figure 6 that each curve is limited to the right since there is a 
maximum value of V consistent with 
each given gap. 

In fact, if we plot the current versus the repulsion for given bands, it
decreases from $I_0$ to $I_1$
by turning on the repulsion. But this repulsion also increases 
strongly the gap. And the noninteracting system having this new gap has
a current much
lower than $I_1$. This is why the effect of the interaction for a fixed gap 
favours the tunneling of electrons through the molecule.

\subsubsection{Influence of the molecule's length.}

The results of the previous subsection can be extended to M=8
and M=10.
As mentioned in the Introduction, 
the current decreases exponentially with the length of the
molecule in the non-interacting case. We find that for various parameters
(repulsion, bandwidth, dimerisation), this 
exponential behavior still holds for interacting
particules, as shown on Figure 7. 

If we choose a set of parameters and consider the
equivalent, non-interacting system (same gap, same bandwidth), this system 
turns out
to depend only very slightly on the size. Therefore, the corresponding current
decreases exponentially with $M$. And we find that the coefficient of this
exponential is larger than in the
interacting case.
(On Figure 7 it is given by the slope of the lines.)
In other words, the current is not only larger when interactions are present 
for a
given size, but the exponential decrease with respect to the length of the
molecule
is slower in the interacting case as
well. This is very important since it is this exponential decrease that puts
limitations on the use of long molecules as wires.

\section{Conclusions}

In this paper, our goal was to estimate the tunneling current through an
interacting oligomere molecule, like polyene. 
The interaction turned out to play an important role.
In this molecule, the charge gap can be reproduced
by a dimerization and no interaction. 
This would correspond to a band insulator in the limit of an infinite molecule.
This gap can also be reproduced by interactions and a smaller dimerization. 
Then we
tend toward the Mott-Hubbard gap limit.

If we fix the band parameters, the charge gap
increases when we turn on the interaction, and our numerical simulations show of
course a decrease of the current.
But for a given bandwidth, if we reduce the dimerization while we increase the
interaction, in order to keep a constant charge gap, then the current grows.
Thus a Mott-Hubbard gap is less damaging than a pure dimerization 
gap for the current.
Furthermore, in the interacting case, the 
persitent current decreases exponentially with the size of
the molecule, like in the non-interacting equivalent case, but more slowly. 
So, the above mentioned difference concerning the current
between a Mott-Hubbard gap and a pure dimerization gap
increases with the size of the molecule.

The next step will be to study
the effect of the spin to check whether our results 
will hold for a realistic molecule
with spin degrees of freedom. 
Clearly, they can already be applied to compounds where
the on-site repulsion is strong compared to the other energy scales,  
and polyene is not so far from this situation, since $t_1 \simeq 2.5 eV$,  
$U \simeq 11.5 eV$ and $V \simeq 2.4 eV$.
Whether this remains true for more general systems is left for future work.

We acknowledge very useful discussions with C. Bruder, C. Joachim and 
M. Magoga.
We are especially indebted to
C. Stafford for very useful explanations concerning persistent currents,
and E. Sorensen for his help with the DMRG. 
The numerical simulations have been performed on the C94 and
C98 of the IDRIS.



\begin{figure}[ht]
\vbox to 283bp {
\centerline{\hbox to 407bp {\includegraphics{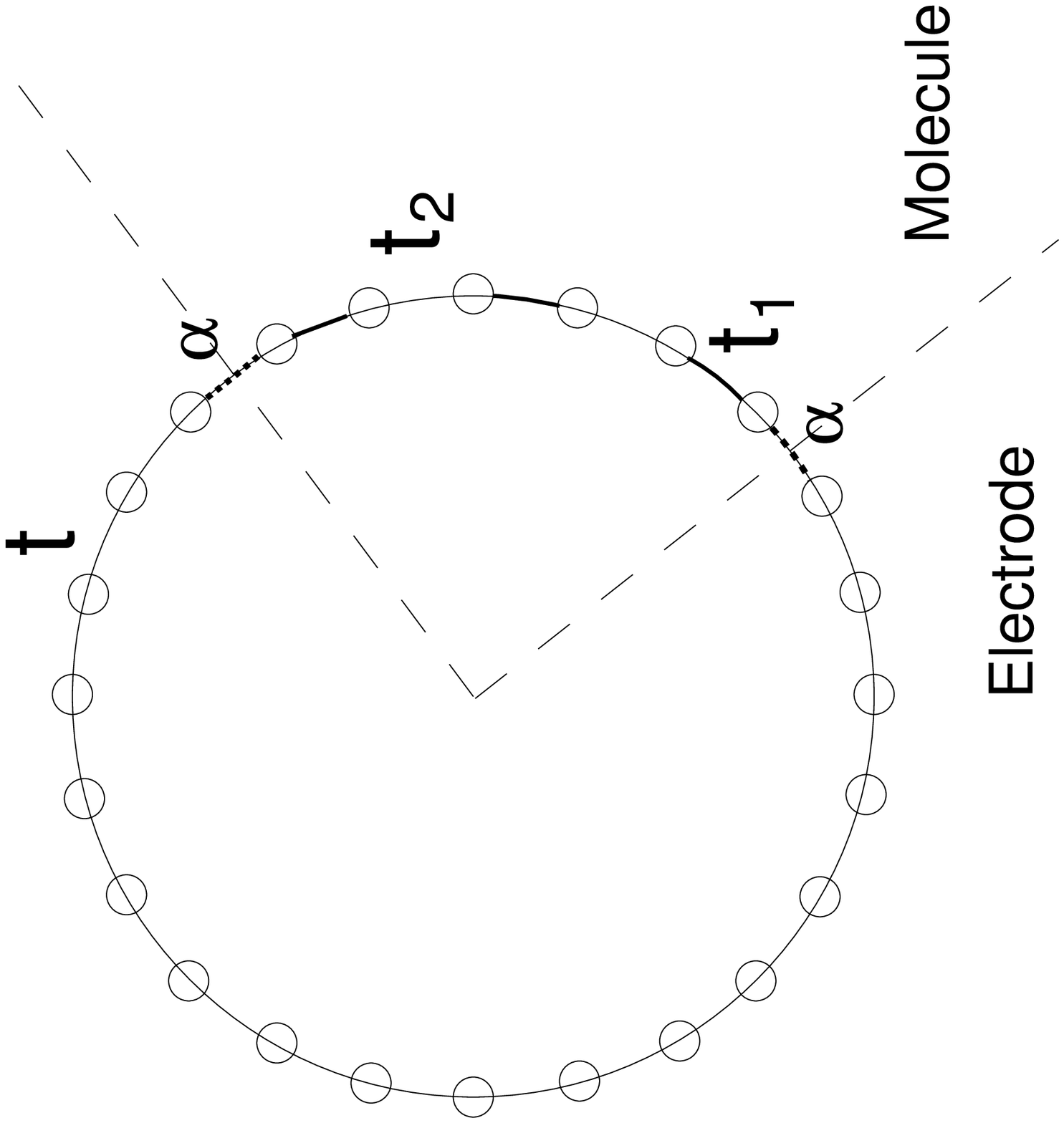}\hfil}}\vfil}
\caption{ Geometry of the system. Each open circle 
corresponds to a site and the links
correspond to the various hopping integrals}
\end{figure}

\begin{figure}[ht]
\vbox to 283bp {
\centerline{\hbox to 407bp {\includegraphics{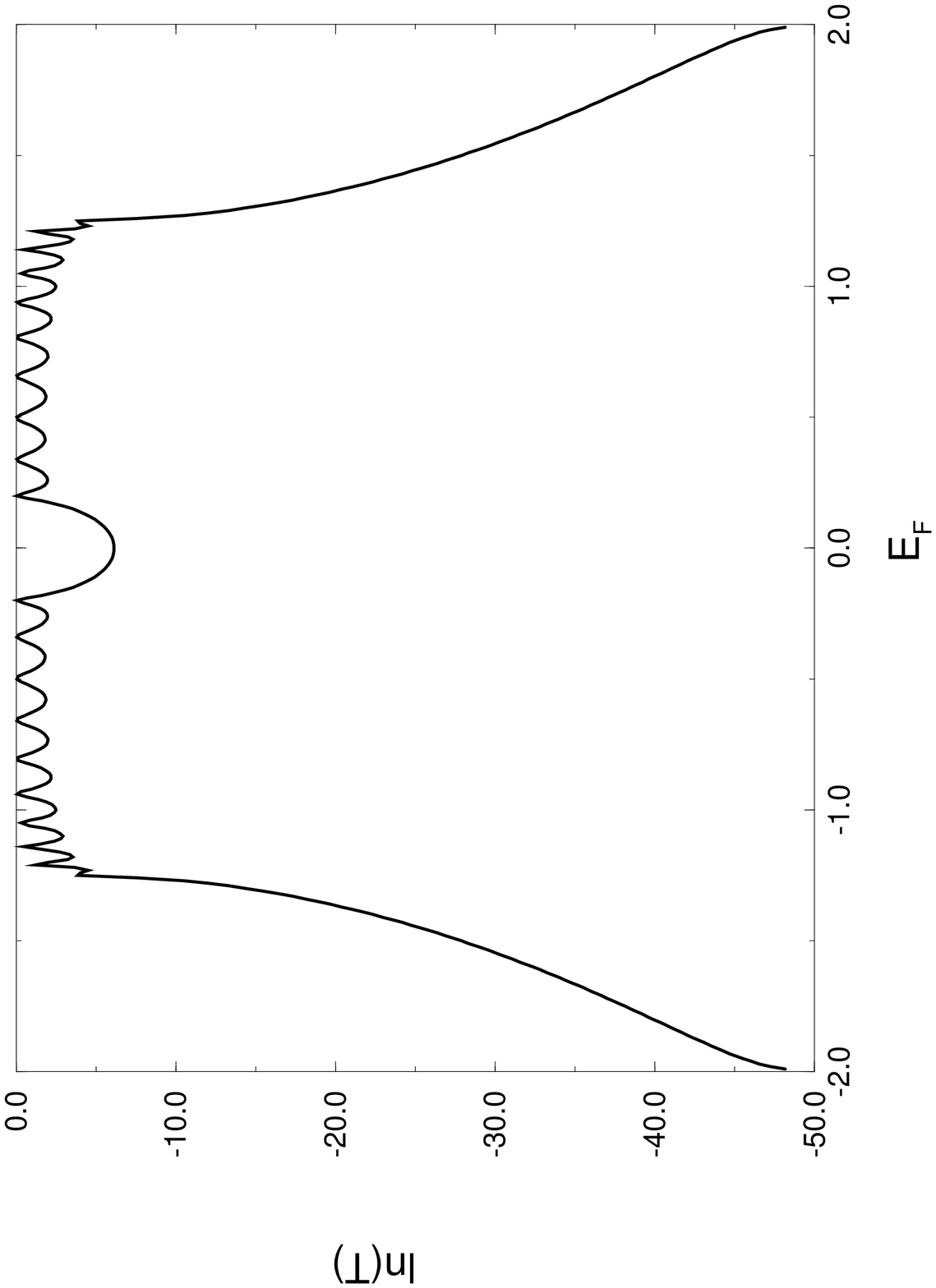}\hfil}}\vfil}
\caption{
Plot of the logarithm of the transmission coefficient T versus the Fermi energy 
$E_F$ of 
the incident particle,
in the non-interacting case. Here $M=20$, $t_1=0.7t$ , $t_2=0.56t$, 
$\alpha=0.35t$, in order to compare this plot to Joachim's result [3].}
\end{figure}

\begin{figure}[ht]
\vbox to 283bp {
\centerline{\hbox to 407bp {\includegraphics{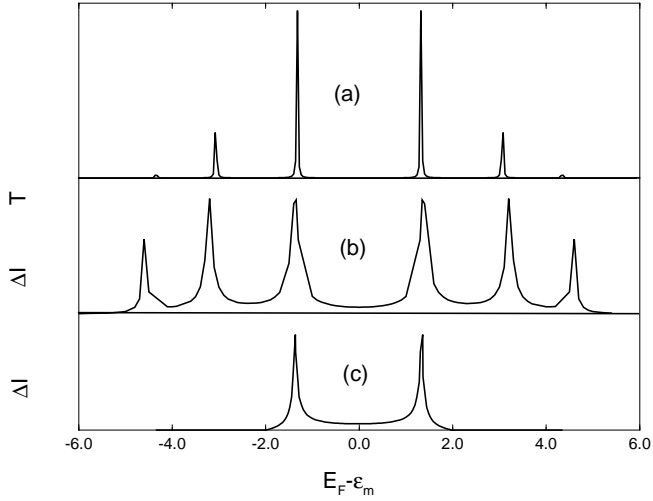}\hfil}}\vfil}
\caption{ Comparison of the transmission coefficient $T$ and the average 
persistent current $\Delta I$ in
the non-interacting case for $M=6$, $t_1=2.6t$ , $t_2=2.2t$, $\alpha=0.3t$. 
(a) $T$ versus the Fermi energy $E_F$. 
The on-site energy in the molecule $\epsilon_m$ was set equal to zero.
(b) Variation of $\Delta I$ 
with $E_F-\epsilon_m$ when $E_F$ is fixed by the band filling
and $\epsilon_m$ varies. The parameters are $N=60$ and $E_F=0$ (half filling).
(c) Variation of $\Delta I$ with $E_F-\epsilon_m$ when $\epsilon_m$ is fixed 
and $E_F$ varies with
the filling. The parameters are $\epsilon_m=0$ and $N=120$.
}
\end{figure}

\begin{figure}[ht]
\vbox to 283bp {
\centerline{\hbox to 407bp {\includegraphics{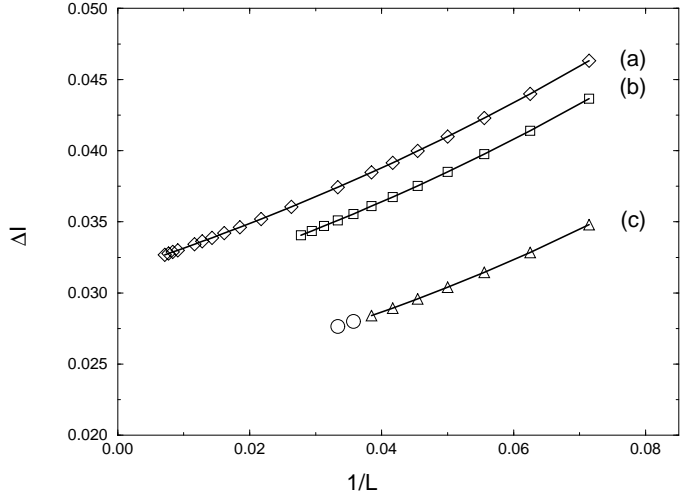}\hfil}}\vfil}
\caption{ Plot of the average persistent 
current $\Delta I$ versus $1/L$.
(a) Non-interacting case up to $L=140$ (diamonds). 
(b) First
order perturbation according to Eq. [6] up to $L=36$ for $V/t=0.2$ (squares).
(c) Exact diagonalizations up to
$L=26$ (triangles) and DMRG for 28 and 30 sites (open circles) 
for $V/t=1$ and $\epsilon_m /t = 0.76$.
In all cases a quadratic fit is correct.
In this example $M=6$, $t_1=2.6t$ , $t_2=2.2t$, $\alpha=0.3t$.}
\end{figure}

\begin{figure}[ht]
\vbox to 283bp {
\centerline{\hbox to 407bp {\includegraphics{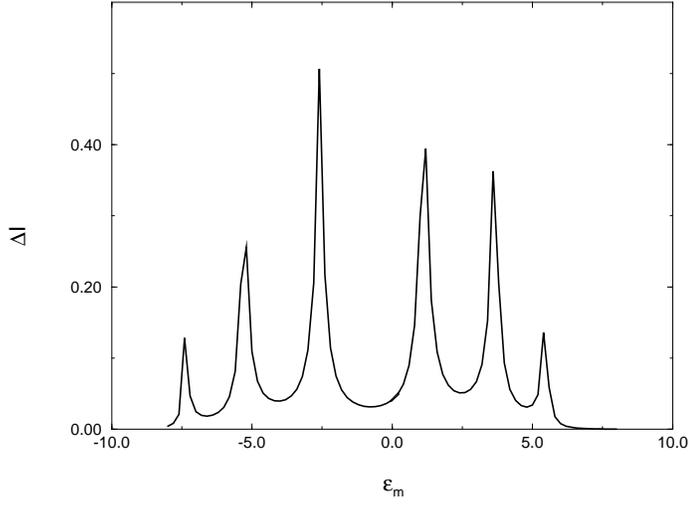}\hfil}}\vfil}
\caption{
Plot of the average persistent current I versus the 
molecular potential in the interacting case for $V/t=1$.
This curve has the same general shape as $I(\epsilon_m)$
in the non-interacting case. Here $M=6$, $L=18$, 
$t_1=2.6t$ , $t_2=2.2t$, $\alpha=0.3t$.}
\end{figure}

\begin{figure}[ht]
\vbox to 283bp {
\centerline{\hbox to 407bp {\includegraphics{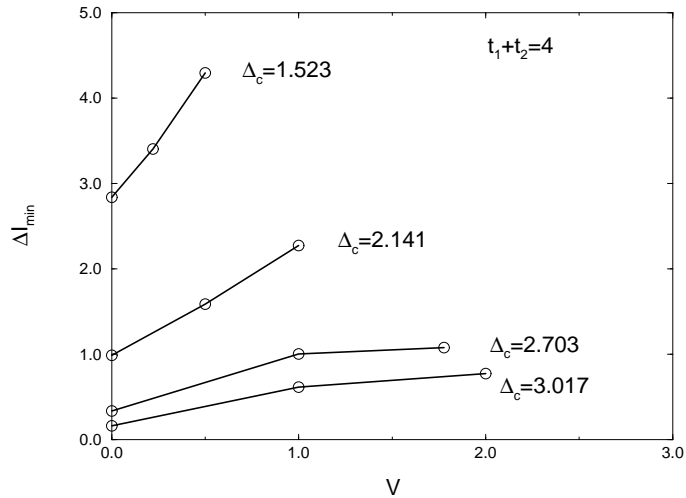}\hfil}}\vfil}
\caption{
Plot of the minimum of the persistent current $\Delta I_{min}$
versus the repulsion V for fixed charge
gap and bandwidth.}
\end{figure}

\begin{figure}[ht]
\vbox to 283bp {
\centerline{\hbox to 407bp {\includegraphics{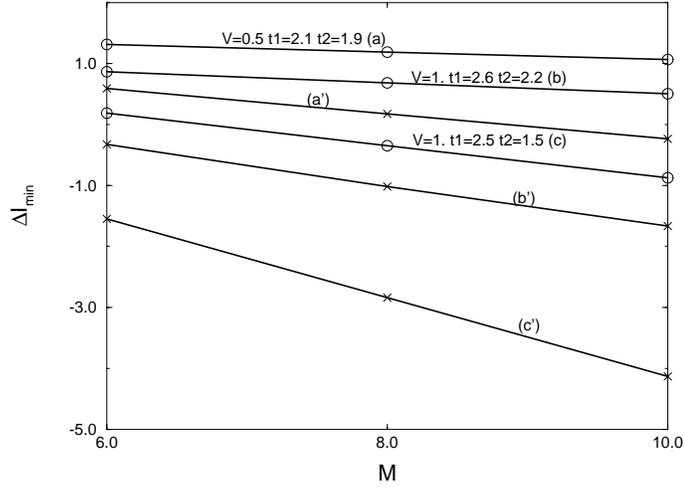}\hfil}}\vfil}
\caption{
Plot of the logarithm of $\Delta I$ versus the size M of the molecule. For each
interacting system (circle) the equivalent non-interacting system (same gap and
band-width) is plotted with crosses. In each example, we get straight lines and 
the curve
with interaction is above the corresponding one without interaction and
decreases more slowly.
}
\end{figure}
\end{document}